\documentclass[aps,twocolumn,superscriptaddress,showpacs,showkeys,floats,
floatats,floatfix]{revtex4}
\usepackage{graphicx,amsmath,amssymb,color}

\begin{document}

\title{Origin of chaos in soft interactions and signatures of non-ergodicity}

\author{M.~W.~Beims$^{1,2}$, C.~Manchein$^1$ and J.~M.~Rost}
\affiliation{Departamento de F\'\i sica, Universidade Federal do Paran\'a,
         81531-990 Curitiba, PR, Brazil}
\affiliation{Max Planck Institute for the Physics of Complex Systems,
         N\"othnitzer Strasse 38, D-01187 Dresden, Germany}
\pacs{05.45.-a,05.45.Ac}
\keywords{Finite-time Lyapunov exponents, chaos, non-ergodicity}
%\date{April, 14th 2004}

\begin{abstract}
The emergence of chaotic motion is discussed for hard, point-like, and
soft collisions between two particles in a one-dimensional box.  It is
known that ergodicity may be obtained in point-like collisions for
specific mass ratios $\gamma=m_2/m_1$ of the two particles, and that
Lyapunov exponents are zero.  However, if a Yukawa interaction between
the particles is introduced, we show analytically that positive
Lyapunov exponents are generated due to double collisions close to the
walls.  While the largest finite-time
Lyapunov exponent changes smoothly with $\gamma$, the number of occurencies 
of the most probable one, extracted from the distribution of finite-time 
Lyapunov exponents over initial conditions, reveals details about the phase 
space dynamics.  In particular the influence of the integrable and
pseudointegrable dynamics without Yukawa interaction for specific mass
ratios can be clearly identified and demonstrates the sensitivity of
the finite-time Lyapunov exponents as a phase space probe.  Being not
restricted to two-dimensional problems such as Poincar\'e sections,
the number of occurencies of the most probable Lyapunov exponents suggest 
itself as a suitable tool
to characterize phase space dynamics in higher dimensions. This is shown
for the problem of two interacting particles in a circular billiard.
\end{abstract}

\maketitle

\section{Introduction}
The investigation of the origin of chaotic motion in standard billiard models
(like the Sinai billiard \cite{sinai}, the Bunimovich stadium \cite{buni} or
the Annular billiard \cite{anular}), has played a pioneering role since the
very beginning of chaos theory. Usually in such models, ballistic
chaotic motion (single particle dynamics) is a consequence of the
spatial billiard geometry.
For interacting many-particles systems, which appear in many areas of
physics, chaotic motion can be generated from the combined effect of
external forces and mutual interactions.  In order to understand how
chaotic motion emerges as a consequence of the interaction
between particles, a simple billiard, namely a one-dimensional box,
will be used.  Since in this case the boundary alone
cannot induces irregular motion of the two particles inside, the role
of the interaction for the generation of chaotic motion becomes clear.

Interacting particles inside billiards can be used to model electrons
in quantum dots.  Electrons are confined inside a disk and are
affected by the surrounding material which composes the semiconductor
\cite{nano}.  In fact, the composition of the surrounding material may
destroy the long Coulomb repulsion between electrons and also change
the effective mass between particles \cite{merkt}.  The influence of
both effects on quantum energy levels and/or electrons dynamics it not
obvious.  However, they can be studied in detail for a physical model
where kinetic and potential energy of the particles can be varied
independently.  This is achieved by a parameter which controls the
range of interaction between the particles (electrons) and by varying
the mass ratio $\gamma=m_{2}/m_{1}$ of the particles.

In this paper we will study the classical dynamics of two interacting particles
inside a one-dimensional billiard as a function of $\gamma$ and  as a
function of interaction range between particles. A Yukawa interaction
between particles is
assumed. Such a system has been considered classically \cite{ulloa}
and quantum mechanically \cite{ulloaQ} for the case of equal masses.
In order to calculate the spectrum of Lyapunov Exponents (LEs), the dynamics
in tangent space is determined explicitly. In the limit of very short range
of interaction, this system should approach the hard-point collision case
analyzed originally by Casati et al \cite{casati}.
Despite ergodic dynamics for the case of
point-like collision at specific $\gamma$, vanishing LEs are a
consequence of the linear instability of this system \cite{zero}.
Such linear unstable systems have become a topical problem in statistical 
mechanics \cite{tsallis} (see also about the origin of diffusion in 
non-chaotic systems \cite{cecconi}).

In the case of a Yukawa interaction, the repulsion between particles
{\it at collisions with the walls} are shown to generate positive LEs.
This is shown explicitly by determining the dynamics in tangent space.
While the mean value of the largest finite-time LE only quantifies the
degree of chaoticity of the system, the number of occurencies of the most 
probable LE, extracted from the distribution over initial conditions, is 
shown to give significant information about regular structures and 
sticky (or trapped) \cite{Zas} trajectories in phase space.  This 
distribution is determined numerically as a function of $\gamma$. 

The plan of the paper is as follows.  Section \ref{1Dpoint} reviews
the main results from the problem of two hard point particles in a
one-dimensional box. In section \ref{1Dyukawa} chaotic motion emerges
with the introduction of the soft Yukawa interaction between the particles.
Positive LEs can be generated from analytical expressions obtained for the
dynamics in tangent space. The distribution of the
largest finite-time LE calculated over the phase space of initial conditions
is used to reveal the underlying dynamics.
Section \ref{1Dyukawa} discusses
the distribution of the largest finite-time LE for the case of two 
interacting particles in a circular billiard. The paper ends with
conclusions in section \ref{conclusion}.

\section{Two particles in a 1D-Box with hard point-like collisions}
\label{1Dpoint}

Two particles in a 1D-box with hard point-like collisions, also called
two-particle hard point gas, can also be treated as a particular case of
the motion of three particles on a finite ring \cite{triangle,triangle2},
which can be mapped onto the motion of a particle in a triangle billiard
\cite{2dtriangle}. In such systems the Lyapunov exponent is zero \cite{zero} 
and the whole dynamics can be
monitored by changing the angles of the triangle billiard. These angles are
functions of the masses ratio between particles. Such triangle billiards have
also been applied to study energy diffusion in one-dimensional systems
\cite{zero}. Although the connection to the triangle billiard is very usefull
to gain insights about collision properties of the particles, it is not
needed for the purpose of the present work.

As follows we summarize the main results obtained by Casati \cite{casati},
which are similar to those observed in the triangle billiard. Using
Poincar\'e sections they \cite{casati} showed that the dynamics is non-ergodic
if $\theta$ is a rational multiple of $\pi$, where

\begin{equation}
\cos{(\theta)}=\frac{1-m_2/m_1}{1+m_2/m_1}=\frac{1-\gamma}{1+\gamma}=\Delta\,.
\label{cost}
\end{equation}
More specifically, writing $\theta=\frac{m}{n}\pi$, where $m$ and $n$
are integers, at most $4n$ distinct velocity values occur.  On the
other hand, when $\theta$ is an irrational multiple of $\pi$,
the velocities become uniformly dense \cite{arnold} in velocity space.
As a consequence, it is at least {\it possible} for the two-particle
hard point gas to be ergodic in velocity space if $\theta/\pi$ is
irrational.  Although Casati and Ford \cite{casati} did not show
explicit results for irrational multiples of $\pi$, they argued that
their numerical results provide evidence supporting ergodic behavior
for irrational $\theta/\pi$ by demonstrating that an increasing number
of velocities is observed for a sequence of rational $\theta/\pi$
approaching an irrational $\theta/\pi$-value.  Note however, that for
irrational $\theta/\pi$ infinite time may be required to observe all
velocities.

Although there are infinitely many mass ratios which give rational
values of $\theta/\pi$, some of them are special.  First, the
integrable cases \cite{Koslov} $\gamma=1,3$ (or $1/3$), which have
$\theta=\frac{1}{2}\pi$ and $\theta=\frac{2}{3}\pi$ (or $\pi/3$),
respectively.  Relating Eq.~(\ref{cost}) with results for the triangle
billiard \cite{gorin} (or even to the polygonal billiard
\cite{berry,genus,gutkin}) it is possible to show that for the
integrable cases the genus is equal $g=1$ (the invariant surface of
the billiard flow is a torus).  For all other rational $\theta/\pi$
the dynamics is pseudointegrable \cite{triangle2} and the invariant
flow is not a torus ($1\le g<\infty$).  For genus $g=2$, the possible
values of $\theta$ are \cite{gorin}: $\frac{1}{5}\pi,\frac{2}{5}\pi$
and the mass ratios are $\gamma\sim0.1$ and $\gamma\sim1.9$,
respectively.  As the genus increases, the invariant surface gets more
and more complicated.  Therefore, besides the integrable cases, the
third special case which has a more ``simpler'' invariant surface,
is expected
for the pseudointegrable case $\gamma\sim1.9$ (we do not use
$\gamma\sim0.1$ because we are interested in values of $\gamma$ in the
interval [$1,4$]).  Later we will come back to the special values
$\gamma=1.0,1.9,3.0$.

The momentum distribution looks quite different for irrational mass ratios.
Although the motion can be ergodic,
the number of momenta increases {\it very} slowly with longer system
evolution \cite{luz}. However, this aspect is not in our present focus.  

\section{Two particles in a 1D-Box with Yukawa interaction}
\label{1Dyukawa}

It is adequate to introduce the center-of-mass and relative coordinates

\begin{equation}
 R=\frac{m_1q_1+m_2q_2}{M},\quad\mbox{and}\quad r=q_1-q_2,
\label{CM}
\end{equation}
respectively, with the total mass $M=m_1+m_2$ and the reduced mass
$\mu=m_1m_2/(m_1+m_2)$.  In these new coordinates, the equations of
motion describe a single composite particle in the hyperspace
($r,R$), called a hyperbilliard \cite{ulloaQ}. The Hamiltonian in
relative coordinates is given by

\begin{equation}
H =  {{P}^2\over 2M} + {{p}^2\over 2\mu} +V_0 {e^{-\alpha r}\over r} = E,
\label{H}
\end{equation}
 where the Yukawa potential $V(r)$ has strength $V_0$ and the parameter
 $\alpha\ge0$ gives the interaction range $r_0=1/\alpha$.
Using scaled coordinates defined by ($\alpha\ne0$)

\begin{eqnarray}
r&=&r_0\tilde r,\quad p=\tilde p\,\sqrt{E},\quad
R=r_0\tilde R, \cr
& &\cr
P&=&\tilde P\,\sqrt{E},\quad
dt=\frac{r_0}{\sqrt{E}}d\tau,
\noindent
\label{CMR}
\nonumber
\end{eqnarray}
and dividing Eq.~(\ref{H}) by $E$, the scaled new Hamiltonian is

\begin{equation}
\tilde H =  {{\tilde P}^2\over 2M} + {{\tilde p}^2\over 2\mu} +
\tilde V(\tilde r)=\epsilon =1,
\label{HS}
\end{equation}
with
\begin{equation}
\tilde V(\tilde r)=\tilde V_0 {e^{-\tilde r}\over \tilde r},\quad
\mbox{and}\quad \tilde V_0={V_0\over r_0E},
\label{V}
\end{equation}
%  still free
and scaled energy $\epsilon =1$. When $\alpha=0$ ($r_0\rightarrow\infty$) the
transformation is  independent of $r_0$ and $\tilde V_0={V_0\over E}$.
Under $\tilde V(\tilde r)$ the composite's particle relative motion is
subject to the force $\tilde Q(\tilde r)=-\partial \tilde V/\partial \tilde r$
while its center of mass motion in $\tilde R$ is free. For the case of equal
masses the chaotic motion of (\ref{H}) was already analyzed \cite{ulloa}.

\subsection{Dynamics in tangent space, Lyapunov exponents (LE),  and the origin of
chaotic motion }
\label{Plyapunov}

This section is dedicated to the analytical calculation of the Lyapunov
spectrum. LEs are very usefull to describe the dynamics in complex
systems \cite{Report}.
When the motion is chaotic, at least one LE is positive. Its
value is determined through the dynamics in tangent space, as will be
shown below.

Between collisions with the walls, equations of motion have the form
\begin{equation}
\tilde F(\tilde \gamma)=(\dot{\tilde r}, \dot{\tilde R}, \dot{\tilde v},
\dot{\tilde V})^{t}= (\tilde v,\tilde V,\tilde Q(r),0)^{t},
% \begin{gather}
% \bf F(\Gamma)=
% \begin{pmatrix} \dot r \\ \dot R \\ \dot v \\ \dot V\end{pmatrix}
%  = \begin{pmatrix} v \\ V \\ Q(r) \\ 0 \end{pmatrix}\,.
\label{F}
% \end{gather}$P(\Lambda_{t}^{p},\gamma)
\end{equation}
and it is easy to see that  center of mass momentum $M\dot{\tilde R}$ is a
constant of motion. In relative coordinates, the composite particle moves
under the influence of the force  $\tilde Q(\tilde r)$.
This is a one-dimensional motion which is regular and integrable.
Collision with left and right walls cause a breaking of the translational
symmetry of the system, and as a consequence, center of mass momentum is 
not a constant of motion anymore.  The effect of left (right) wall
collisions lead to the following change in the  phase space point
$\tilde\gamma_{i} = (\tilde r^{i},\tilde R^{i},\tilde v^{i},\tilde V^{i})$
before the collision to
$\tilde\gamma_f = (\tilde r^{f},\tilde R^{f},\tilde v^{f},\tilde V^{f})$
 after the collision, $\tilde \gamma_{f}={\tilde{\bf D}}_{k}\tilde\gamma_{i}$,
with
\begin{gather}
%\begin{pmatrix} r^f \\ R^f \\ v^f \\ V^f \end{pmatrix} =(-)^k
\tilde{\bf D}_{k}=(-1)^k
    \begin{pmatrix} 1 & \quad 0 & \quad 0 & \quad 0  \\
                    0 & \quad 1 & \quad 0 &  \quad 0 \\
                    0 & \quad 0 & \quad -\Delta & \quad 2  \\
                    0 & \quad 0 & \quad 2\frac{\mu}{M} & \quad \Delta
\end{pmatrix}\,,
%\begin{pmatrix} r^i \\ R^i \\ v^i \\ V^i \end{pmatrix},
\label{left}
\end{gather}
where $\Delta=(m_1-m_2)/M$.  The label $k=1(2)$ is used for particle
$1(2)$.  Since the two particles can never pass each other in the
one-dimensional billiard, it is assumed without loss of generality
that particle $1$($2$) never collides with the right(left) wall.  The
complete time evolution in phase space can be formulated by
integrating the equation of motion between collisions with the walls
and by taking into account $\tilde\gamma_{f}$ from Eq.~(\ref{left})
each time the composite particle collides with the walls.

In order to calculate Lyapunov exponents, the time evolution of an
infinitesimal  path difference (the nearby trajectory) in the scaled
tangent space ($\delta\tilde\gamma$) must be determined, given by
\begin{equation}
\delta{\tilde \gamma}(\tau)= 
\tilde{\mathbf M}(\tau)\delta{\tilde\gamma}(\tau_0),
\label{mgamma}
\end{equation}
with the scaled monodromy matrix
\begin{equation}
\tilde {\mathbf M}(\tau)=\frac{d{\tilde\gamma}(\tau)}{d{\tilde\gamma}(\tau_0)}.
\label{M}
\end{equation}
Lyapunov exponents are the average rates of growth or shrinkage of
such infinitesimal changes that are the eigenvectors of $\tilde{\bf M}$,
\begin{equation}
\tilde\lambda_i=\lim_{\tau\rightarrow\infty} \frac{\log{\tilde\mu_i(\tau)}}
{\tau}\,,
\label{lyap}
\end{equation}
where $\tilde\mu_i(\tau)$ is the $i$-th eigenvalue of ${\bf \tilde M}$.  The
matrix $\tilde{\bf M}$ can be written itself as a product of matrices for
small time steps.  Since the motion between collisions is regular, Lyapunov
exponents are zero.
The situation is different for collisions of the {composite particle}
with the walls. We follow the algorithm developed by Dellago et
at \cite{dellago} to formulate the equations in the scaled tangent space 
according to

\begin{equation}
\delta\tilde\gamma_f= \frac{\partial \tilde C}{\partial\tilde\gamma_{i}}\delta
\tilde\gamma_i
+ \left[ \frac{\partial \tilde C}{\partial\tilde\gamma_{i}}\tilde F(\tilde
\gamma_i)- \tilde F(\tilde C(\tilde \gamma_i))\right]\delta\tau_f\,
\label{alg}
\end{equation}
where
$\tilde C=\mathbf{\tilde D}_{k}\tilde\gamma_i$ gives the transformation at
collisions with the walls  with $\mathbf{\tilde D}_{k}$ from 
Eq.~(\ref{left}).
$\delta\tau_{f}$ is the delay in the collision time of the (infinitesimal)
nearby trajectory with respect to the collision time of the reference
trajectory. Between the collision time of the main trajectory and the
collision time of the nearby trajectory, the {composite particle} moves
under the influence of the force  $\tilde Q(r)$ and the delay time can be
determined from

\begin{equation}
t_k=\frac{\partial S_k}{\partial E}=
m_k\int_{q_k^0}^{q_k^n}\frac{\partial \dot{q_k}}{\partial E}dq_k
=\int_{q_k^0}^{q_k^n}\frac{dq_k}{\dot{q_k}}\,,
\label{tk}
\end{equation}
where $S_k$ is the action of the $k$ particle,  $q_k^0$ is the position 
of the nearby trajectory when the main trajectory collides with the wall, 
and $q_k^n$ is the collision point of the nearby trajectory. The energy 
dependence for $\dot q_k$,  
$\dot q_1=\sqrt{2m_1(E-\frac{1}{2}m_2\dot q_2^2-V_0\frac{e^{-\alpha r}}{r})}$,
is obtained from the energy conservation.
Using Eqs.~(\ref{CM}) we have

\begin{eqnarray}
q_{1(2)}&=&R-\frac{(-)^{1(2)}m_{2(1)}}{M}r,\\
& & \cr
\dot q_{1(2)}&=&V-\frac{(-)^{1(2)}m_{2(1)}}{M}v,
\end{eqnarray}
and Eq.~(\ref{tk}) can be written as 

\begin{eqnarray}
t_k&=&\int_{R_0}^{R_n}\frac{M d R}{M V-(-)^{1(2)}
  m_{2(1)} v}\cr
 & & \cr
 & & \cr
&-&(-)^{1(2)}
  \int_{r_0}^{r_n}\frac{m_{2(1)}d r}{M
V-(-)^{1(2)}m_{2(1)} v}.
\label{tk2}
\end{eqnarray}
Since $V$ and $v$ does not depend on $R$, the integral in $R$
can be determined analytically. After the integration of Eq.~(\ref{tk2}), 
terms proportional to $\delta R^i= R_0-R_n$ and  $\delta r^i= r_0-r_n$ will
appear (quadratic terms $(\delta R)^2$ and $(\delta r^i)^2$ can be neglected 
in the 
linear analysis) and it can be written in scaled relative coordinates as

\begin{equation}
-\delta\tau_{f}=\tilde t_{k}=\frac{t}{\sqrt{E}}=
\tilde A_k\delta \tilde R^i+\tilde B_k\delta \tilde r^i,
\label{mtk}
\end{equation}
where
\begin{eqnarray}
\tilde A_{1(2)}&=&\frac{M}{M\tilde V^i-(-)^{1(2)}
  m_{2(1)}\tilde v^i},\cr
& & \cr
\tilde B_{1(2)}&=&\frac{-(-)^{1(2)}}{\delta \tilde r^i}
  \int_{\tilde r_0}^{\tilde r_n}\frac{m_{2(1)}d\tilde r}{M\tilde
V-(-)^{1(2)}m_{2(1)}\tilde v}.\cr
\label{AB}
\nonumber
\end{eqnarray}
Finally, using Eqs.~(\ref{F}),(\ref{left}) and (\ref{mtk}), in (\ref{alg}),
the collision of the {composite particle} in tangent space  with
left ($k=1$) and right walls ($k=2$) under a force
$\tilde Q(\tilde r)=-\partial \tilde V/\partial \tilde r$ takes the form
$\delta\tilde \gamma_{f}=\mathbf{\tilde M}_{k}\delta\tilde\gamma_{i}$ with
%\begin{gather}
    \begin{equation}
%\delta \bf\Gamma_f
	\mathbf{\tilde M}_{k}=(-)^k
\begin{pmatrix}
    -\Delta & \quad 2 & \quad 0 &\quad 0  \\
 & & & \\
    2\frac{\mu}{M} & \quad \Delta  & \quad 0 &\quad 0 \\
 & & & \\
 -\Delta_k\tilde Q\tilde B_k  & \quad  -\Delta_k\tilde Q\tilde
   A_k   & \quad -\Delta &
\quad 2  \\
 & & & \\
   - 2\frac{\mu}{M}\tilde Q\tilde B_k &\quad  -2\frac{\mu}{M}\tilde Q \tilde
A_k& \quad
2\frac{\mu}{M}
   & \quad \Delta\,,
\end{pmatrix}
%\delta \bf\Gamma_i,
\label{zp}
\end{equation}
%\end{gather}
where  $\Delta_k=-[\Delta+(-1)^k]$. The determinant of $\mathbf{\tilde M}_{k}$
is equal to $1$ and eigenvalues are also equal to $|1|$. However, the
matrix elements proportional to $\Delta_k,\tilde Q,\tilde A_k$ and
$\tilde B_k$ generate positive LE when the global monodromy matrix
$\bf \tilde M$ is constructed. Therefore, the presence of the interaction
force $\tilde Q$ and terms
$\tilde A_k$ and $\tilde B_k$, related to the time delay in the tangent-space
collision dynamics, are essential for the chaotic properties of the system.
Otherwise, if $\tilde Q=0$, no positive LE can be obtained.

 Two limiting situations can be discussed.  a) The long range
 interaction ($\alpha=0.0, r_0\rightarrow\infty$): at each collision
 with the wall, the interaction force $\tilde Q(r)$ is finite and
 positive LEs are generated; b) The short range interaction
 ($r_0\ll1$): in general $\tilde Q(r)$ is close to zero if particles
 are sufficiently separated and the LEs are zero.  However, the
 dynamics becomes chaotic due to a {\it double collision} process.
 For example, assuming that particles $1$ and $2$ are moving close
 together towards the left wall with a mutual repulsion close to zero
 due to the short range nature of the interaction.  As particle $1$
 collides with the left wall, it changes its direction and moves
 towards particle $2$, interacting with it.  If such a {\it double
 collision} occurs infinitely close to the wall, the terms $\tilde
 Q\tilde B_k$ and $\tilde Q\tilde A_k$ are not necessarily zero, a
 positive LE is generated and chaotic motion appears.  This crucial
 role of double collisions to generate chaotic motion was also
 observed in another model of two interacting particles \cite{luz}.
 The chaotic motion induced by the soft potential will be discussed in
 more detail in the next section.

\subsection{Signatures of regular motion in the distribution of
largest finite-time Lyapunov exponents}
\label{Dlyapunov}

From the description of the dynamics by the monodromy matrices as
constructed in the last section, it is clear that chaotic dynamics is
generated in the presence of a soft interaction potential.  The
interesting question, however, is, if any signatures of non-ergodicity
from the hard-point collision,  or stickiness, 
can be identified.  To this end we have investigated the distribution
$P(\Lambda_{t},\gamma)$ of the finite-time largest Lyapunov exponents
\cite{finite} $\Lambda_{t}$ as a function of the mass ratio $\gamma$.
In general, for infinite time, the LEs $\Lambda_{\infty}$ are well
defined and do not depend on initial conditions.  This holds also true
for reasonably large finite times, if the motion is ergodic and the
Lyapunov spectrum has good convergence properties.  In
quasi-regular systems, however, where the chaotic trajectory may
approach a regular island and can be trapped there for a while, the
value of the local LE can decrease.  This will affect the convergence
of $\Lambda_{t}$ which depends now on the initial conditions.  On the
other hand, it implies that the distribution $P(\Lambda_{t})$,
calculated over many initial conditions, contains information about
the amount of regular motion (and sticky trajectories) in phase
space.  Usually, for fully chaotic systems $P(\Lambda_{t},\gamma)$ has a 
Gaussian distribution (see for example \cite{grebogi} and references therein).
%%%%%%%%%%%%%%%%%%%%%%%%%%%%%%%%%%%%%%%%%%%%%%%%%%%%%%%%%%%%%%%%
 \begin{figure}[htb]
 \unitlength 1mm
 \begin{center}
 \includegraphics*[width=8.5cm,angle=0]{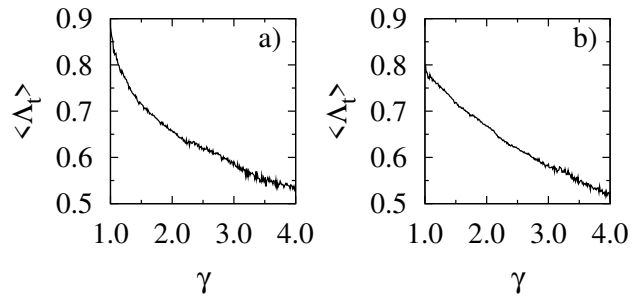}
% \leavevmode
% {\epsfxsize 8.5cm
% \epsfbox{dist.eps}}
 \end{center}
 \caption{Mean value of the finite-time largest Lyapunov exponent
 calculated over $400$ trajectories up to time $t = 10^{4}$ and at scaled
 energy $\epsilon=1$, for (a) long interaction range ($r_0\rightarrow\infty$),
  i.~e.~$\tilde V_{0}=0.1$, and  (b) short range interaction ($r_0=0.1$),
   i.~e.~$\tilde V_{0}=1.0$, with the hard walls located at $q=\pm 1$.
  For each trajectory the largest LE is evaluated over $10^5$ initial 
  conditions.}
  \label{mean}
  \end{figure}
%%%%%%%%%%%%%%%%%%%%%%%%%%%%%%%%%%%%%%%%%%%%%%%%%%%%%%%%%%%%%%%%%%%%%

\begin{figure}[htb]
 \unitlength 1mm
 \begin{center}
 \includegraphics*[width=8.6cm,angle=0]{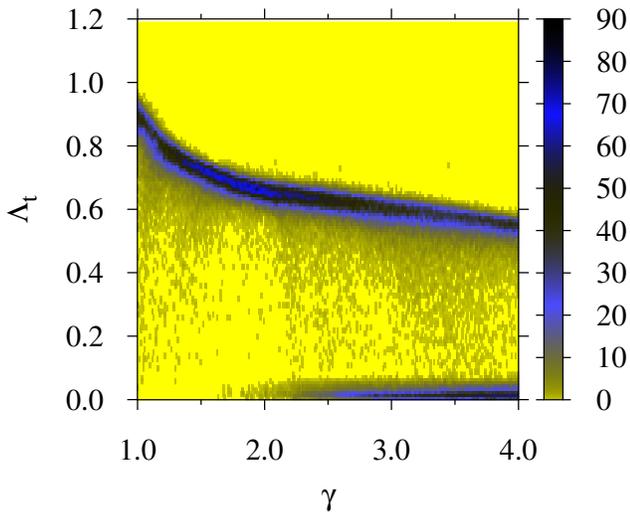}
% \leavevmode
% {\epsfxsize 8.5cm
% \epsfbox{dist.eps}}
 \end{center}
 \caption{(color online). 
Finite-time distribution of the largest Lyapunov exponent
 $P(\Lambda_{t},\gamma)$ calculated over $400$ trajectories up to time
 $t = 10^4$ and for $r_0\rightarrow\infty$.  With increasing
 $P(\Lambda_{t},\gamma)$ the color changes from light to dark (white
 over yellow and blue to black).}
  \label{dist}
  \end{figure}

The mean $\langle\Lambda_{t}\rangle$, shown  in Fig.~\ref{mean}(a)-(b),
decreases monotonically from roughly $0.9$ to $0.54$ in Fig.~\ref{mean}(a)
($r_0\rightarrow\infty$), and from $0.80$ to $0.53$ in Fig.~\ref{mean}(b)
($r_0=0.1$), over a change of $\gamma$ from $1.0$ to $4.0$. This means
that the dynamics is getting more and more regular as expected
since  $\gamma\rightarrow\infty$ constitutes an 
integrable limit with the heavy particle at rest.  Figure \ref{dist} 
shows the finite-time distribution of the largest LE, $P(\Lambda_{t},\gamma)$,
for the case of long range interaction ($r_0\rightarrow\infty$). It
reveals two indications for increasingly regular motion with
growing mass ratio: a) the {\it value} of
$\langle\Lambda_{t}(\gamma)\rangle$ itself decreases and b) an
increasing number of initial conditions lead to
$\Lambda_{t}(\gamma)$ close to zero (related to sticky trajectories),
or converge exactly to zero (regular islands).  At $\gamma=4.0$ for
example, about 15\% of the initial conditions lead to
$\Lambda_{t}=0$.  The gray points below the main curve are
related to chaotic trajectories which were trapped for a while close
to regular islands.  Some examples will be given below.
%%%%%%%%%%%%%%%%%%%%%%%%%%%%%%%%%%%%%%%%%%%%%%%%%%%%%%%%%%%%%%%%
%%%%%%%%%%%%%%%%%%%%%%%%%%%%%%%%%%%%%%%%%%%%%%%%%%%%%%%%%%%%%%%%%%%%
An interesting feature in  Fig.~\ref{dist} is the change of the width
of $P(\Lambda_{t},\gamma)$
around the most probable $\Lambda^{p}_{t}$ defined through
\begin{equation}
    \label{probable}
   \left. \frac{\partial
P(\Lambda_{t},\gamma)}{\partial
\Lambda_{t}}\right|_{\Lambda_{t}=\Lambda_{t}^{p}}\,=0.
\end{equation}
For mass ratio between $\gamma\sim 1.5$ and $\gamma\sim 2.2$, many
initial condition lead to the same $\Lambda_{t}$.  In this region,
$\Lambda^{p}_{t}(\gamma)$ has a maximum as a function of $\gamma$.  In
other words, almost all initial conditions converge to the same LE and
the dynamics should approach an ``ergodic-like'' motion.  In fact, in
this region the gray points (related to sticky or trapped trajectories) below
the main curve almost disappear.

%%%%%%%%%%%%%%%%%%%%%%%%%%%%%%%%%%%%%%%%%%%%%%%%%%%%%%%%%%%%%%%%
 \begin{figure}[htb]
 \unitlength 1mm
 \begin{center}
 \includegraphics*[width=8.6cm,angle=0]{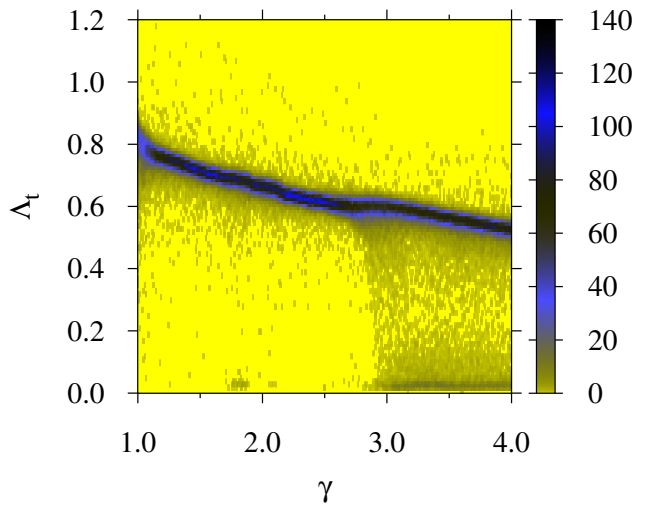}
% \leavevmode
% {\epsfxsize 8.5cm
% \epsfbox{dist.eps}}
 \end{center}
 \caption{(color online). 
Finite-time distribution of the largest Lyapunov exponent
 $P(\Lambda_{t},\gamma)$ calculated over $400$ trajectories up to time
 $t = 10^4$ and for $r_0=0.1$.  With increasing
 $P(\Lambda_{t},\gamma)$ the color changes from light to dark (white
 over yellow and blue to black).}
  \label{dist2}
  \end{figure}
%%%%%%%%%%%%%%%%%%%%%%%%%%%%%%%%%%%%%%%%%%%%%%%%%%%%%%%%%%%%%%%%%%%%
A similar behavior is found in the short interaction limit ($r_0=0.1$)
shown in Fig.~\ref{dist2}.  Compared to Fig.~\ref{dist}, two main
differences can be observed: firstly, the maximum for
$\Lambda^{p}_{t}(\gamma)$ from Fig.~\ref{dist} (between
$\gamma\sim1.5$ and $\gamma\sim2.2$) is divided into two maxima, one
close to $1.5$, and the other one close to $2.4$.  Therefore, a
minimum of $\Lambda^{p}_{t}(\gamma)$ appears in between
($\gamma\sim1.9$), where trapped trajectories are expected to occur
more often if compared with the two maxima $\gamma\sim1.5$ and $2.4$.
Secondly, the abrupt appearance of many gray points below the main
curve close to $\gamma\sim2.7$, which may indicate that a regular
island is born.  This will be shown in more details below.

\begin{figure}[htb]
 \unitlength 1mm
 \begin{center}
\includegraphics*[width=8.5cm,angle=0]{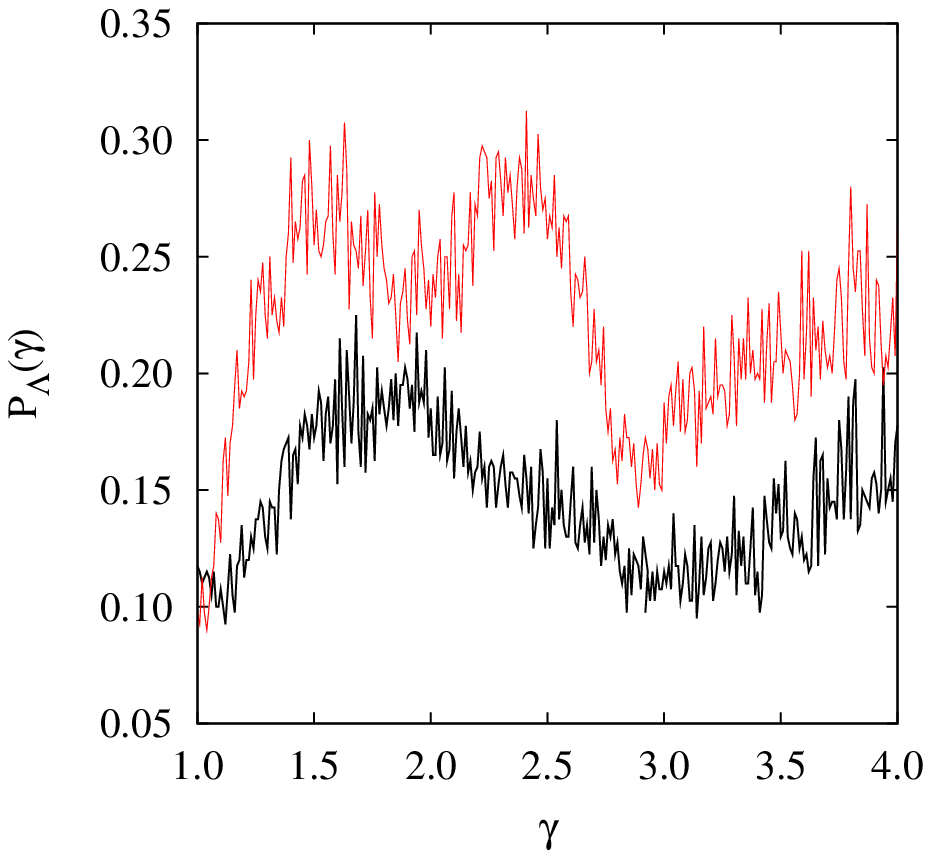}
\includegraphics*[width=8.5cm,angle=0]{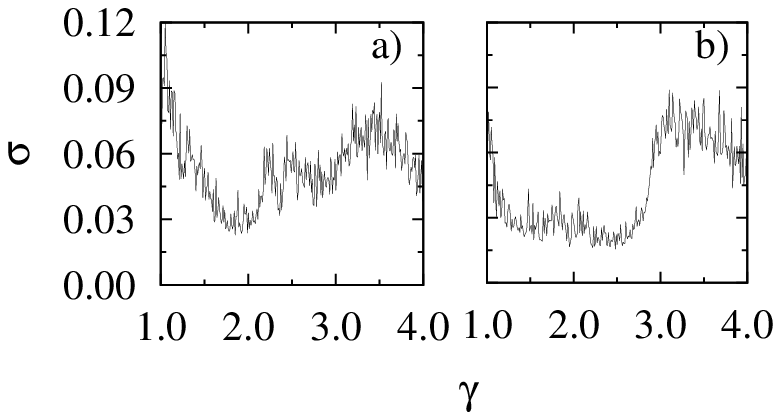}
% \leavevmode
% {\epsfxsize 8.5cm
% \epsfbox{dist.eps}}
 \end{center}
\caption{Top: Normalized distribution $P_{\Lambda}(\gamma)$ of the most
probable Lyapunov exponent $\Lambda_{t}^{p}$ for $r_0\rightarrow\infty$ 
 (black) and $r_0=0.1$ (gray). Bottom: $\sigma$ for a) $r_0\rightarrow\infty$ 
 and b) $r_0=0.1$}
  \label{max}
  \end{figure}

A more systematic way to uncover this trend is to follow
$P(\Lambda_{t}^{p},\gamma)\equiv P_{\Lambda}(\gamma)$ as a function of the
mass ratio $\gamma$ shown in Fig.~\ref{max} (top). If $P_{\Lambda}$ is large, a
large fraction of initial conditions lead to the same $\Lambda_{t}$ and trapped
trajectories are rare. For example, the maximum of $P_{\Lambda}(\gamma)$ in
Fig.~\ref{max} (top, $r_0\rightarrow\infty$, black curve) close to 
$\gamma\sim1.9$, is the region in Fig.~\ref{dist} where gray points below the 
main curve are rare.
%%%%%%%%%%%%%%%%%%%%%%%%%%%%%%%%%%%%%%%%%%%%%%%%%%%%%%%%%%%%%%%%
%%%%%%%%%%%%%%%%%%%%%%%%%%%%%%%%%%%%%%%%%%%%%%%%%%%%%%%%%%%%%%%%%%%%%
The fast variation of $P_{\Lambda}(\gamma)$ is due to statistical
fluctuations in the determination of $\Lambda_{t}$ over the 400
initial conditions.  However, two main valleys can be identified in
the black curve of Fig.~\ref{max}, one close to $\gamma=1$ and the
other close to $\gamma=3$.  These are exactly the mass ratios for
which the hard point-like collision dynamics (section \ref{1Dpoint})
is integrable.  The gray curve of Fig.~\ref{max} for the short
interaction ($r_0=0.1$), which is closer to the limit of hard-point
collisions, presents an additional little valley at $\gamma\sim1.9$.
This is the pseudointegrable case with genus $g=2$ which appears in
the hard-point collision.  In other words, if $\gamma$ is close to
values for which the dynamics in the point-like gas is integrable ($g=1$) 
or ``simpler'' ($g=2$), then the dispersion around $\Lambda^{p}_{t}$ 
increases so that $P_{\Lambda}(\gamma)$ decreases, and signatures
of non-ergodicity 
are expected under
additional Yukawa interaction.  Another interesting observation is
that the minimum at $\gamma\sim 1.9$ (see gray curve from
Fig.~\ref{max}) disappears in the long interaction limit
$r_0\rightarrow\infty$ (see black curve from Fig.~\ref{max}).  It
means that the regular motion from the integrable cases of the
hard-point collision survives longer under the perturbation of the
soft interaction than the regular motion from the pseudointegrable
case.

For fully chaotic systems the quantity $P_{\Lambda}(\gamma)$ is just
the maximum of a Gaussian distribution and it should increases
linearly with $t$, since the variance
$\sigma=(\langle \Lambda_t^2 - \langle \Lambda_t\rangle^2\rangle)^{1/2}$ 
for such systems goes with $1/t$. This behavior of $\sigma$ has been 
observed by studying ergodicity in high-dimensional sympletic maps 
\cite{Falcioni}, and used for the 
detection of small islands in the standard map \cite{Tomsovic}.  For our 
case we found that the time dependence of  $P_{\Lambda}(\gamma)$ goes with
$\sim t^{0.3}$, which means that a significant number of islands is present. 
In order to compare $\sigma$ with 
$P_{\Lambda}(\gamma)$, Figs.~\ref{max}a) and b) show the behavior of 
$\sigma$ as a function of $\gamma$. Each time trapped trajectories are
present, $\sigma$ should has a maximum, exactly the opposite behavior 
from $P_{\Lambda}(\gamma)$. In Fig.~\ref{max} a) we see two maxima, one
close to $\gamma=1.0$ which corresponds to the minima of $P_{\Lambda}(\gamma)$
for the same value of $\gamma$, and the other one at $\gamma\sim3.5$ which as 
no analogous for $P_{\Lambda}(\gamma)$. For Fig.~\ref{max} b) we again see two 
maxima, one close to $\gamma=1.0$ and the other one located between
$\gamma\sim3.0$ and $\gamma\sim3.5$. Besides the peaks at $\gamma=1.0$,
the other peaks are not located at the integrable ($\gamma=3.0$)
and pseudo-integrable ($\gamma\sim1.9$) cases from the hard point-like 
collisions cases. The peak around  $\gamma\sim3.0$ in  Fig.~\ref{max} b) is 
correct, but by far not precise. This shows that the quantity $\sigma$ is not 
sensible enough to detect signatures of the regular dynamics from the 
hard-point collision case. The reason for this is that $\sigma$ is not 
able to detect small regular islands which appear in the phase space, as for 
example the points related to zero LEs in Fig.~\ref{dist2} for $\gamma=1.9$.
It is worth to mention that the maxima and minima from $P_{\Lambda}(\gamma)$ 
does not change significantly with the number of initial conditions.

We can conclude that signatures from regular structures in phase-space
exist and are uncovered by the dispersion of the largest Lyapunov exponent
$\Lambda_{t}$ most clearly visible along the cut $P_{\Lambda}(\gamma)$ defined
by the number of occurrences of the most probable Lyapunov exponent 
$\Lambda_{t}^{p}$ (Eq.~\ref{probable}). No such signatures, however, are 
visible in less sensitive quantities such as the mean Lyapunov exponent 
$\langle\Lambda_{t}\rangle$ or the fluctuation $\sigma$.

%%%%%%%%%%%%%%%%%%%%%%%%%%%%%%%%%%%%%%%%%%%%%%%%%%%%%%%%%%%%%%%%%%%%%
 \begin{figure}[htb]
 \unitlength 1mm
 \begin{center}
 \includegraphics*[width=4.cm,angle=0]{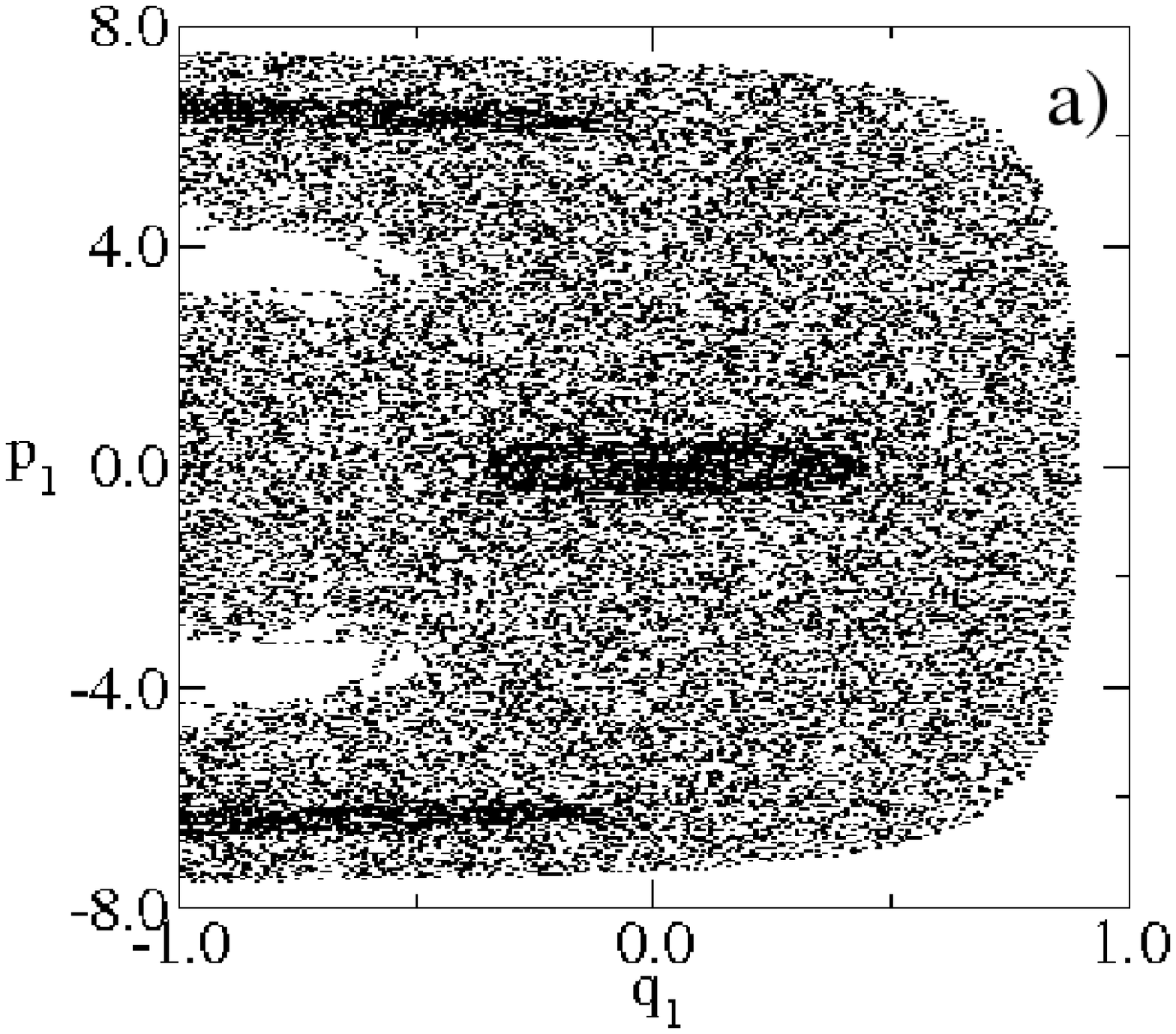}
 \includegraphics*[width=4.cm,angle=0]{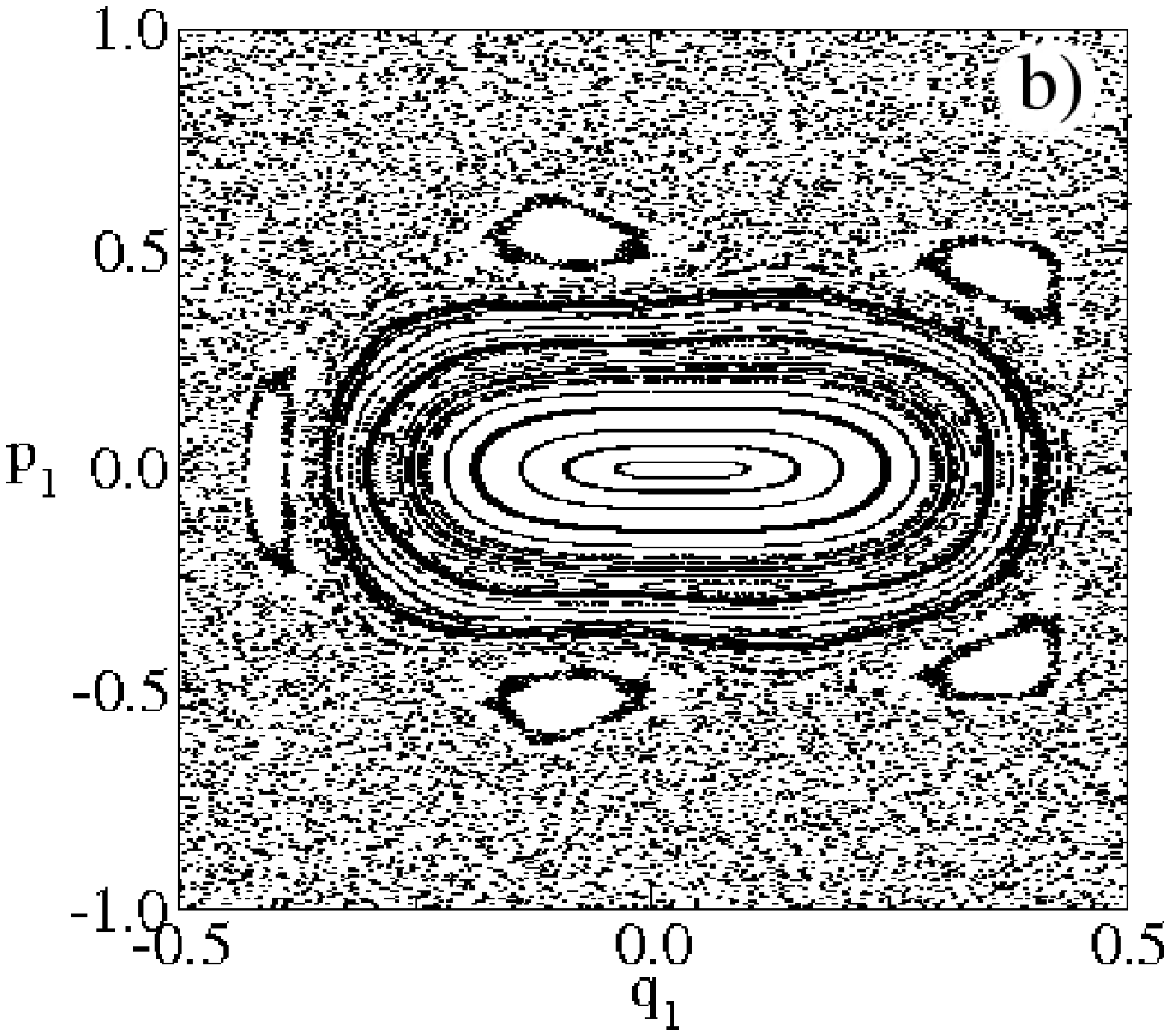}
 \includegraphics*[width=4.cm,angle=0]{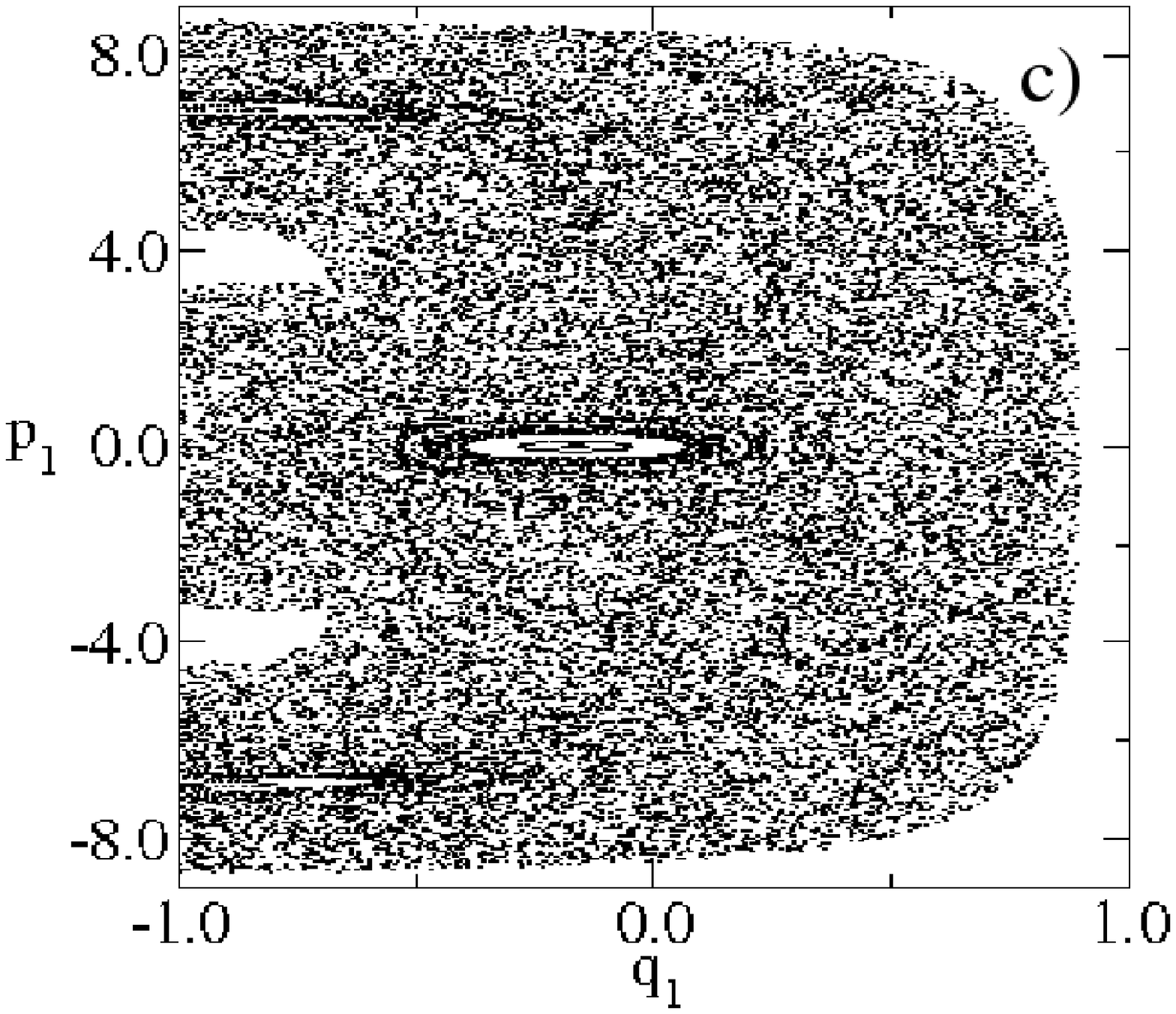}
 \includegraphics*[width=4.cm,angle=0]{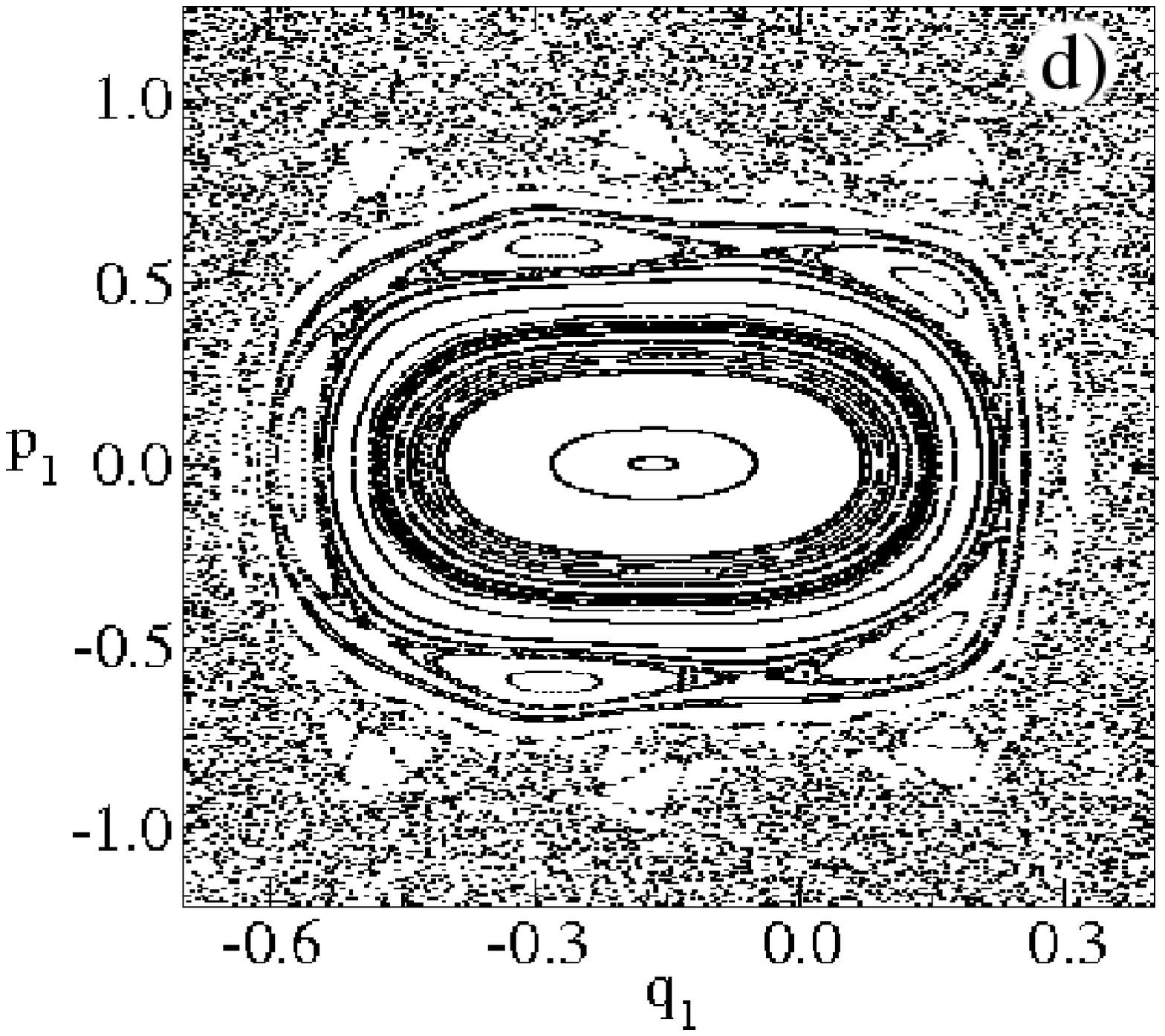}
 \end{center}
 \caption{Poincar\'e Surfaces of Section for $r_0\rightarrow\infty$ and 
    a) $\gamma=3.0$ and  c) $\gamma=4.0$. Figures c) and d) are the 
      corresponding magnifications of the quasi-regular regions.}
  \label{PSS}
  \end{figure}
%%%%%%%%%%%%%%%%%%%%%%%%%%%%%%%%%%%%%%%%%%%%%%%%%%%%%%%%%%%%%%%%%%%%%
 The interpretation of the Lyapunov properties are supported by the
 relevant PSS. Figure \ref{PSS} (top) shows PSSs for interaction
 $r_0\rightarrow\infty$ and cases $\gamma=3.0$ and $\gamma=4.0$, where
 $P_{\Lambda}(\gamma)$ has a minimum and maximum (see black curve in
 Fig.~\ref{max}), respectively.  Although the system is more chaotic
 for $\gamma=3.0$ than for $\gamma=4.0$, trapped trajectories appear
 near the island for $\gamma=3.0$, see the corresponding magnification
 in Fig.~\ref{PSS} (top, right).  Such trapped motion near regular
 islands, which does not appear for $\gamma=4.0$ (see
 Fig.~\ref{PSS}(bottom, right)), affects $P(\Lambda_{t}^{p},\gamma)$
 and consequently, $P_{\Lambda}(\gamma)$ has a minimum near
 $\gamma=3.0$.
%%%%%%%%%%%%%%%%%%%%%%%%%%%%%%%%%%%%%%%%%%%%%%%%%%%%%%%%%%%%%%%%%%%%%
 \begin{figure}[htb]
 \unitlength 1mm
 \begin{center}
 \includegraphics*[width=4.25cm,angle=0]{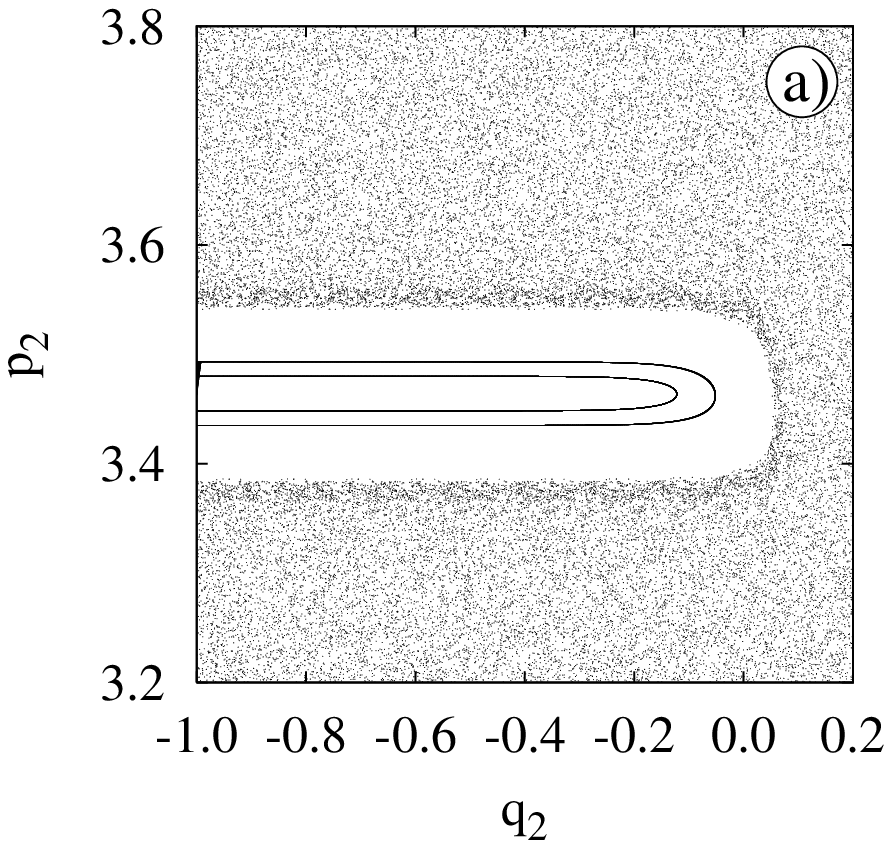}
 \includegraphics*[width=4.25cm,angle=0]{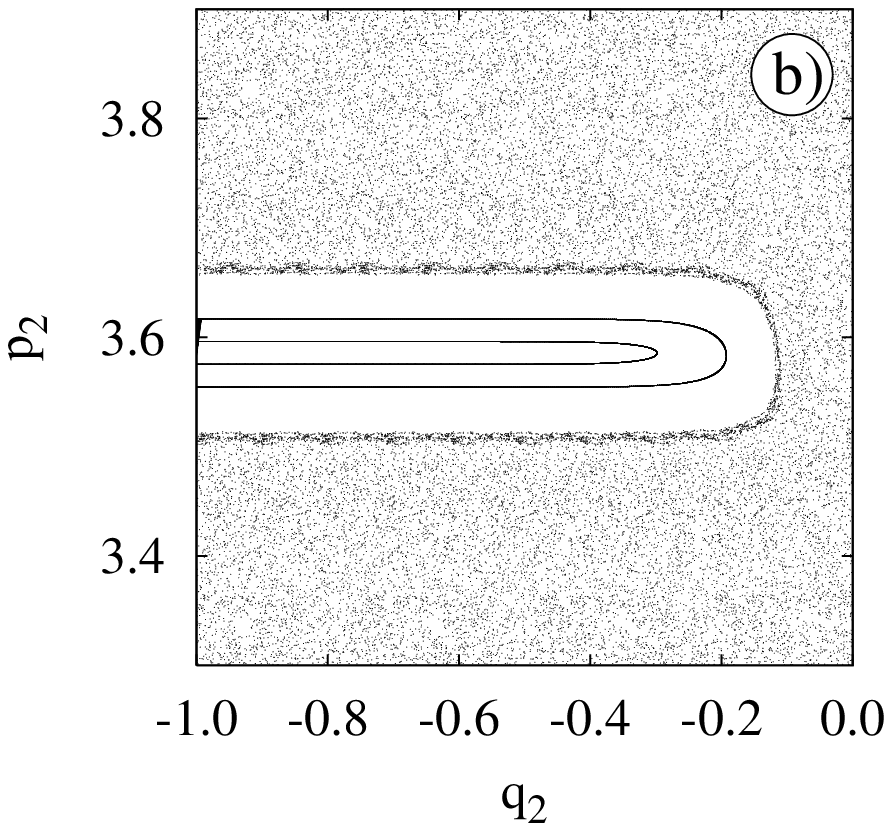}
 \includegraphics*[width=4.25cm,angle=0]{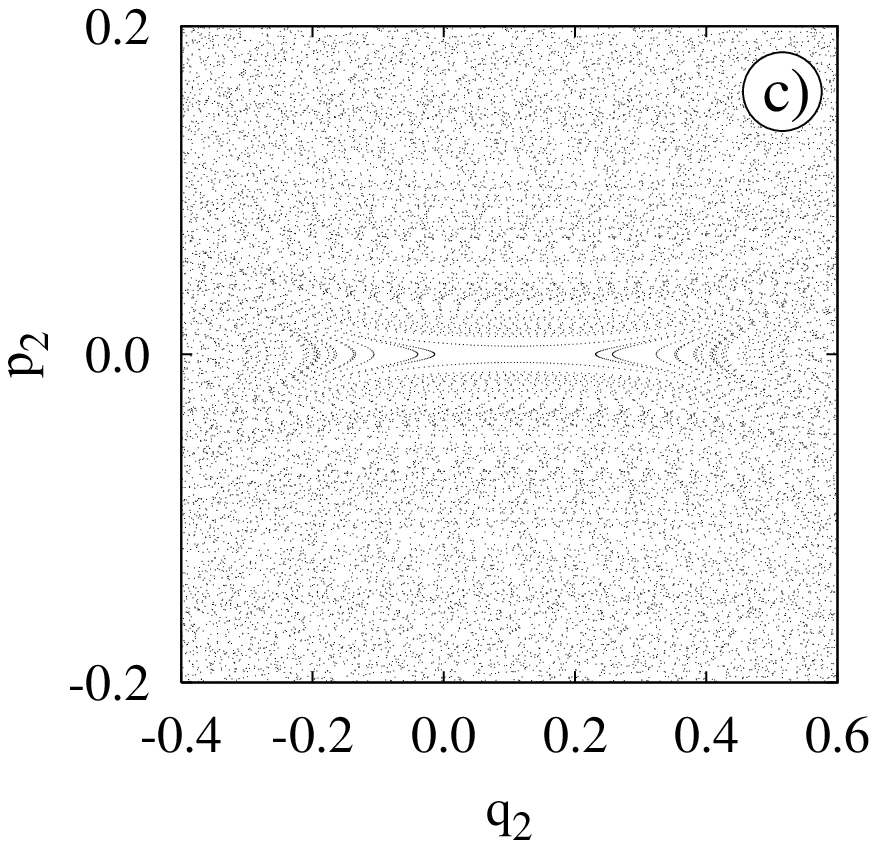}
 \includegraphics*[width=4.25cm,angle=0]{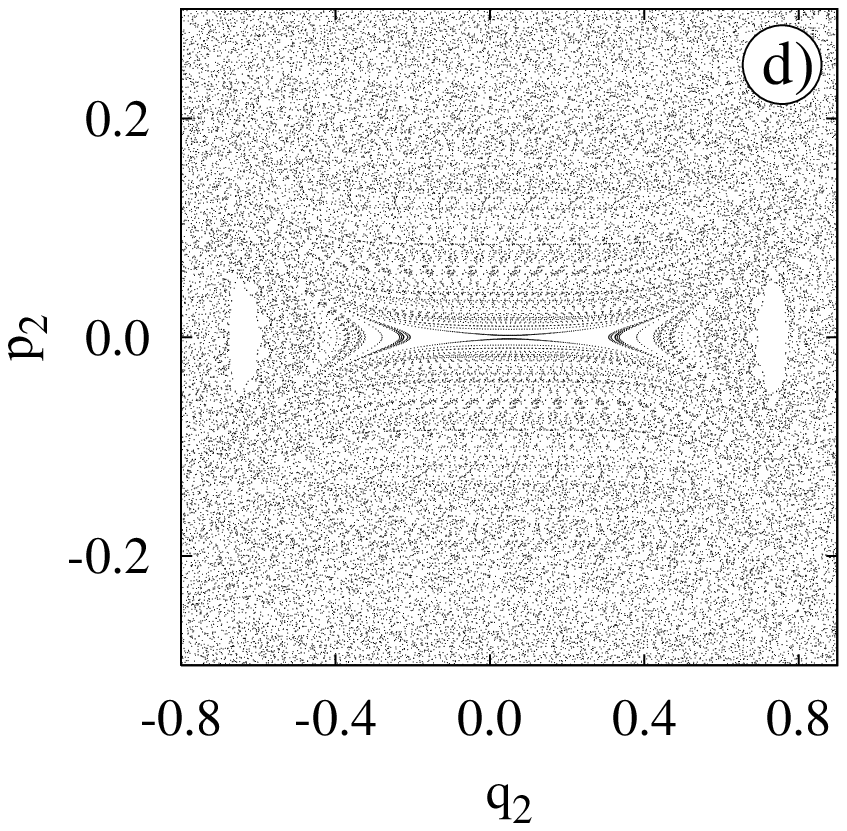}
 \end{center}
 \caption{Magnification of the Poincar\'e Surfaces of Section with
   $r_0=0.1$ for a) $\gamma=1.5$, b) $\gamma=1.8$, c) $\gamma=2.6$ and
   d) $\gamma=2.8$.}
  \label{PSS2}
  \end{figure}
%%%%%%%%%%%%%%%%%%%%%%%%%%%%%%%%%%%%%%%%%%%%%%%%%%%%%%%%%%%%%%%%%%%%%
Another example is shown in Fig.~\ref{PSS2} for $r_0=0.1$.  For
$\gamma=1.5$ (top, left), which has a maximum in Fig.~\ref{max} (top, gray
curve), no trapped trajectories are found up to the time propagated.  For
$\gamma=1.8$ (top,right) which has a minimum in Fig.~\ref{max} (top, gray
curve), trapped trajectories start to appear.  Moreover, the abrupt 
appearance of gray points for $\gamma\sim2.7$ below the main curve
Fig.~\ref{dist2} can be nicely explained using the PSS. Figure
\ref{PSS2} compares the PSSs for $\gamma=2.6$ (bottom, left) with
$\gamma=2.8$ (bottom, right), where gray points appear in
Fig.~\ref{dist2}.  Clearly, it can be seen that when a regular island
is born trapped trajectories start to appear around the island and, as
a consequence, many initial conditions with lower LE are obtained
(gray points in Fig.~\ref{max}).  The regular island is related to a
period-4 orbit which has the following property: for each second hit
of particle 1 with the left wall, particle 2 is at rest.  There is a
similar periodic orbit for the hard-point collision case
\cite{gutkin}.

Note that although $P_{\Lambda}(\gamma)$ is determined only from
trajectories with positive $\Lambda_{t}$, it provides information
about the amount of regular structure in phase space through chaotic
trajectories which are trapped close to regular islands.  Since
trapped trajectories are characteristic for mixing in phase space,
$P_{\Lambda}(\gamma)$ provides a tool to analyze phase space mixing.

\section{Two particles in a Circular Billiard with Yukawa interaction}
\label{2Dyukawa}

In order to show the utility of $P_{\Lambda}(\gamma)$ for systems with higher 
dimensions, we discuss now the case of two interacting particles in a 
circular billiard. The interaction between particles is still of Yukawa type.
The chaotic motion is now generated by the combined effect of the curvature
of the walls from the circular billiard and the double collisions discussed
in last section. The phase space is 8-dimensional and it is not possible to 
construct an adequate PSS which allows to look at the underline dynamics.
Trajectories will fill out any chosen PSS and no information about details
of the dynamics can be
obtained. Sticky trajectories, for example, which may cause non-ergodicity 
due to a partial focusing of trajectories \cite{donnay}, are difficult to
detect. In such partial focusing, an infinitesimal family of nearby 
trajectories that starts out parallel will lead to Lyapunov exponents which 
converge very slowly in time. This is a typical behavior in high 
dimensional quasi-regular systems. In this section we show the effectiveness
of $P_{\Lambda}(\gamma)$ to obtain relevant informations in high-dimensional 
quasi-regular systems. 

\begin{figure}[htb]
 \unitlength 1mm
 \begin{center}
 \includegraphics*[width=4.6cm,angle=0]{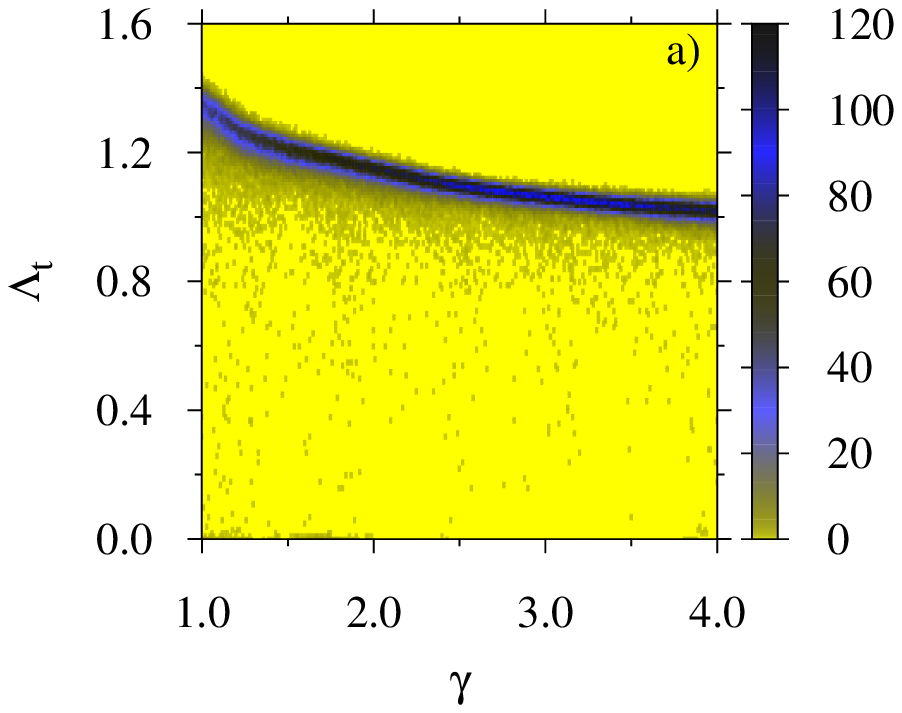}
 \includegraphics*[width=3.8cm,angle=0]{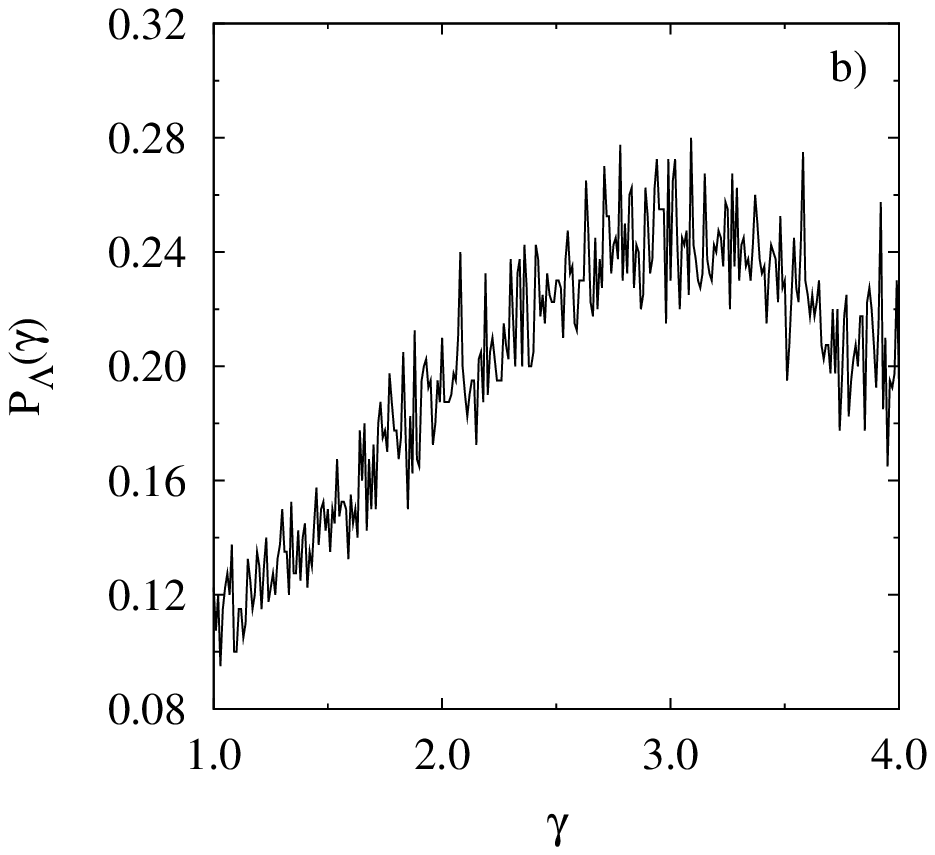}
%vspace{-1cm}
% \includegraphics*[width=3cm,angle=0]{fig7c.eps}
% \leavevmode
% {\epsfxsize 8.5cm
% \epsfbox{dist.eps}}
 \end{center}
 \caption{(color online). 
  a) Finite-time distribution of the largest Lyapunov exponent
 $P(\Lambda_{t},\gamma)$ calculated over $400$ trajectories up to time
 $t = 10^4$ and for $r_0\rightarrow\infty$ for the circular billiard. 
 With increasing
 $P(\Lambda_{t},\gamma)$ the color changes from light to dark (white
 over yellow and blue to black); b) Normalized distribution 
$P_{\Lambda}(\gamma)$ of the number of occurrencies of the most probable 
Lyapunov exponent $\Lambda_{t}^{p}$.}
  \label{dist2D}
  \end{figure}

Figure \ref{dist2D}a) shows the finite-time distribution of the largest LE
for the case of long range interaction ($r_0\rightarrow\infty$) in the 
circular billiard. The value of the the mean LE calculated of the $400$ 
initial conditions, decreases growing the mass 
ratio and the regular motion increases. This is the only information we can 
get from the LE about the complicated dynamics of the two interacting 
particles inside the billiard.
%%%%%%%%%%%%%%%%%%%%%%%%%%%%%%%%%%%%%%%%%%%%%%%%%%%%%%%%%%%%%%%%
%%%%%%%%%%%%%%%%%%%%%%%%%%%%%%%%%%%%%%%%%%%%%%%%%%%%%%%%%%%%%%%%%%%%
However, the gray points below the main curve of Fig.~\ref{dist2D}a) are 
related to chaotic 
trajectories which were trapped for a while close to regular islands.  
Therefore some additional information about sticky trajectories and 
the amount of quasi-regular motion in phase space may be obtained from 
$P_{\Lambda}(\gamma)$. This is shown in Fig.~\ref{dist2D}b).
One minima is observed close to $\gamma\sim1.0$ where 
trapped trajectories and partial focusing of trajectories in phase
space are expected. On the other hand, a more ergodic-like motion is expected 
for $\gamma\sim3.0$, where $P_{\Lambda}(\gamma)$ has a maximum. This is
different from the results for the 1D-Box from last section (see the 
minimum for $r\rightarrow\infty$ at $\gamma=3.0$ in Fig.~\ref{max} Top).  
In other words, we clearly see that signatures from the integrable case 
$\gamma=3.0$ from the 1D hard point-like collision case disappear in the two 
dimensional case. This tells us that the dynamics from the integrable case 
$\gamma=3.0$ is a consequence that particles
cannot pass each other in the 1D case. As the masses ratio increase to 
$\gamma=4.0$, the LE decreases and the system is more regular. In this
case $P_{\Lambda}(\gamma)$ also decreases slowly and the amount of
regular islands, and the motion around them, increases. 

\section{Conclusions}
\label{conclusion}

Usually, chaotic motion is generated through non-linear equations of motion. 
As a consequence, exponential divergence of nearby trajectories occurs 
which can be quantified by positive Lyapunov exponents. Another source of 
chaotic motion emerges even for linear equations of motion through  boundary
conditions. The advantage of such systems is the possibility to obtain
rigorous mathematical results. One example is a point particle moving
among high-dimensional cylindrical scatterers \cite{astrid}, which is
similar to the high-dimensional Lorentz gas. In these systems, strongly
chaotic motion is generated due to the convex curvature of hard disks or 
spheres. In fact, the collision time delay between nearby trajectories
due to the curvature of the surface, is responsible for the chaotic motion.

Curved boundaries are not present in the one-dimensional confinement considered in this paper.  For the hard point-like collision case non-ergodic motion is generated when the mass ratio gives a $\theta$ value [from Eq.~({\ref{cost})] which is a rational multiple of $\pi$. Ergodic motion, on the other hand, may be obtained for irrational multiples of $\theta/\pi$.  We have shown that for additional soft Yukawa interaction between the two particles, chaotic motion is obtained for any mass ratio.  {\it Double} collisions of particles, which occur very close to the walls, are essential to generate positive LEs in the short interaction range limit.  The collisional time delay in tangent space, together with the soft Yukawa interaction, are responsible for the chaotic motion.  The mean of the largest finite-time LE $\langle\Lambda_{t}\rangle$ decreases smoothly as the mass ratio increases and does not provide detailed information about the phase space structure.  This type of information is provided by the probability distribution of the largest finite-time Lyapunov exponent $P(\Lambda_{t},\gamma)$.  It reveals that the dispersion around $\langle\Lambda_{t}\rangle$ increases when trapped trajectories are present in the phase space.  We have shown that a cut through $P(\Lambda_{t},\gamma)$ along the number of occurrencies of the most  probable Lyapunov exponent,  $P_{\Lambda}(\gamma)$, gives a quantitative measure of the influence of regular motion in mixed phase space. Specifically for the system studied here, we have shown that  $P_{\Lambda}(\gamma)$ decreases when the structure in phase-space is more regular and the mass ratio is close to the integrable cases $\gamma=1,3$ (genus $g=1$) or to the ``simpler'' dynamics (pseudointegrable at $\gamma\sim1.9$, with $g=2$) from the hard point-like collision.  We also observed that the regular motion of the integrable cases from the hard-point collision survives longer under the perturbation of the soft interaction than the regular motion from the pseudointegrable case.  Hence, the dynamics under the additional Yukawa interaction, although in principle chaotic, ``remembers'' the integrable and pseudointegrable dynamics in the system without the soft Yukawa interaction.  This is certainly a subtle effect and therefore, we expect that in general the number of occurrencies of the most probable Lyapunov exponent provides a sensitive tool to probe details in phase space dynamics. We have also shown that this quantity is much more sensitive if compared with the mean square fluctuations of the LE. In contrast to Poincar\'e sections this tool is easily applicable in higher dimensional systems, where trapped trajectories may cause non-ergodicity due to a partial focusing of trajectories \cite{donnay}. In order to show this we calculated $P(\Lambda_{t},\gamma)$ for two interacting particles in a circular billiard, where the phase space is $8$-dimensional and it is not possible the construct an adequate Poincar\'e section to analyze the dynamics. We show that by increasing the mass ratio, the mean LE decreases and the system gets more regular. Furthermore, for a minimum of $P_{\Lambda}(\gamma)$ at $\gamma\sim1.0$ trapped trajectories are expected, and a maximum at $\gamma\sim3.0$ ergodic-like motion is expected. Therefore, $P_{\Lambda}(\gamma)$ can be used in higher-dimensional systems as a tool to characterize the dynamics.

\vspace*{1cm}
\section*{Acknowledgments}
\noindent
CM thanks CAPES and MWB thanks CNPq for financial support. MWB is grateful
to M.~G.~E.~da Luz, J.~D.~Szezech, A.~S.~de Wijn and S.~Tomsovic for helpful 
discussions.

%\end{multicols}
\end{document}